\documentstyle[preprint,aps,epsfig]{revtex}
\begin{document}
\tightenlines
\draft
\title{The $a_0(980)$, $a_0(1450)$ and $K_0^*(1430)$ Scalar Decay Constants \\
and the Isovector Scalar Spectrum}
\author{Kim Maltman} 
\address{Department of Mathematics and Statistics, York University, \\
          4700 Keele St., Toronto, Ontario, CANADA M3J 1P3 \\ and}
\address{Special Research Centre for the Subatomic Structure of Matter, \\
          University of Adelaide, Australia 5005}
\maketitle
\begin{abstract}
The scalar
correlator $\Pi (q^2)=i\int d^4x\, e^{iq\cdot x}
\langle 0\vert T\left( J(x)J^\dagger (0)\right)\vert 0\rangle$ (with
$J=\partial_\mu\left(\bar{d}\gamma^\mu u\right)$) is studied using
a class of finite energy sum rules shown recently to be very well satisfied
in the vector isovector channel.  The values 
of the $a_0$ scalar decay constants extracted in 
this analysis, which describe the
couplings of the $a_0(980)$ and $a_0(1450)$ to the current
$J$, are shown to be comparable, strongly disfavoring any scenario in
which the $a_0(980)$ is interpreted as
a loosely bound $K\bar{K}$ molecule and the $a_0(1450)$ is assigned
to the same flavor multiplet as the $K_0^*(1430)$.
The $a_0$
decay constants are also compared to the analogous decay constant
describing the coupling of the $K_0^*(1430)$
to the divergence of the strange
vector current, $J_s=\partial_\mu\left( \bar{s}\gamma^\mu u\right)$
(which may be obtained from experimental $K_{e3}$ and $K\pi$ phase
shift data) and implications of the relative magnitudes for the
interpretation of the nature of the $a_0$ states discussed.
\end{abstract}
\pacs{14.40.Cs,12.39.Mk,11.55.Hx,11.40.-q}
\section{Introduction}
Despite considerable recent theoretical activity, there exists,
at present, no concensus on the nature of the
$f_0(980)$ and $a_0(980)$ mesons.  Interpretations still advocated in the
literature include (1) the $q^2\bar{q}^2$ (``four-quark'') cryptoexotic
interpretation originally proposed by Jaffe\cite{jaffe} (see
Refs.~\cite{mnachasov,nnachasov}); (2) the $K\bar{K}$ bound state,
or ``molecule'', picture\cite{kkwi,kkspeth,kkkl,kkoo,kklmz}; (3) the
unitarized quark model picture\cite{tornqvist,morgan,mp,zou,uqmas,kklmz};
and (4) the ``minion'' picture of Gribov\cite{gribov}.

The $K\bar{K}$ molecule assignment for the
$f_0(980)$ and $a_0(980)$ is attractive (and, perhaps for this reason, 
often cited as already established) because it naturally
explains the proximity to $K\bar{K}$ threshold and strong $K\bar{K}$
couplings of the states.  Such bound states occur naturally
in the quark model treatment of Ref.~\cite{kkwi}, as well as several
coupled channel treatments using model meson-meson interactions
with parameters fit to 
experimental data\cite{kkspeth,kkkl,kkoo,kklmz}.
The picture predicts $\gamma\gamma$ decay
widths\cite{barnes,lcb91,kkoo}
compatible with experiment\cite{pdg98,bp98}, and 
considerably smaller than those expected, at least in the simplest version
of the quark model, for ${}^3P_0$
$q\bar{q}$ mesons.
One should bear in mind, however, that an alternate quark model approach,
whose $\gamma\gamma$ model dynamics are constrained
by an analogous treatment of $\pi^0,\eta\rightarrow\gamma\gamma$, allows
widths compatible with experiment for
a conventional quark model assignment\cite{aabn99}, and that
small $\gamma\gamma$ widths are also claimed in the four-quark
picture, based on a schematic estimate loosely motivated by the MIT bag
model\cite{ads82}.  An alternate means of testing
the molecule scenario is via the processes $\phi\rightarrow f_0\gamma\ ,
a_0\gamma$\cite{cik93}.  Recent experimental determinations\cite{f0g,a0g}
give branching ratios larger than predicted in the
molecule picture\cite{cik93,ags97} but, 
since the predictions of Refs.~\cite{cik93}
and \cite{ags97} differ by a factor of 4, 
realistic theoretical uncertainties may be large, making
the discrepancy with experiment difficult to interpret reliably.

One of the problems with the interpretation of the light
scalar states is that their
narrow experimental widths, which appear 
unnatural for a conventional $q\bar{q}$ meson assignment
(but natural in the $K\bar{K}$ molecule picture),
may not, in fact, reflect the true intrinsic widths, since 
strong coupling of an intrinsically
broad state to a nearby $s$-wave threshold can produce 
significant narrowing through the effects of channel coupling
and unitarity\cite{flatte}.  Given the empirically observed
strong $f_0(980)$ and $a_0(980)$ couplings to
$K\bar{K}$, the possibility arises that these states are,
not molecules, but rather ``unitarized quark model'' (UQM) states,
i.e. conventional $q\bar{q}$ mesons with properties
strongly distorted by
coupling to the nearby $K\bar{K}$ threshold\cite{tornqvist}.
In such a scenario, one expects significant $K\bar{K}$
content in the scalar states, and hence most likely a significant
reduction in the $\gamma\gamma$ widths\cite{tornqvist} (though 
these widths have yet to be calculated in the UQM picture).
The distortion caused by the strong coupling will presumably also alter
expectations for $\phi\rightarrow f_0\gamma ,\ a_0\gamma$, though one 
again awaits an explicit calculation of this effect.

A number of observations
made in the context of the UQM, and discussions surrounding it,
bear repeating at this point.  First, in the presence of strong coupling to
nearby decay channels, a single underlying state can manifest itself
as more than just a single pole close to the physical region
(T\"ornqvist and Roos\cite{tornqvist}, for example, claim that
both the $a_0(980)$ and $a_0(1450)$ are likely to be manifestations
of the same underlying $q\bar{q}$ state).  Second, existing
experimental data are apparently insufficient to distinguish 
between the molecule
and UQM pictures, since coupled 
channel models exist, both with and without an underlying
conventional quark model state, which provide good fits to existing
data in the coupled $\pi\pi$, $K\bar{K}$ and $\pi\eta$, $K\bar{K}$ channels,
once the data has been used to fit the free parameters of the 
models. Third,
although
the method for distinguishing conventional resonances from
bound states posited by Morgan and Pennington\cite{mp} (a conventional
resonance having poles near the physical region
on both the second and third sheets,
a bound state having such a pole on only the second sheet) is
frequently borne out in existing coupled channel 
calculations\cite{kkwi,kkspeth,kkkl,kkoo,kklmz}, 
this is not universally the case\cite{kklmz}.  In particular,
a situation in which there are two nearby poles,
one on the second and one on the third sheet, can arise in a model
which provides a good fit to existing data, but for which
one of the nearby poles is most naturally thought of as
the coupled-channel remnant of a $K\bar{K}$
bound state (in the sense that, as the channel coupling
is dialed up towards its final fitted value, what was a 
$K\bar{K}$ bound state in the absence of channel coupling
moves continuously to become one of the two final nearby poles)\cite{kklmz}.
Thus, even if data could distinguish between the one-nearby-pole 
and two-nearby-pole scenarios (see Refs.~\cite{mp,kkspeth,basz} for
arguments that this may not be possible at present),
it appears that it would not allow us to distinguish between
the UQM and molecule scenarios.


In this paper, we discuss the
feasibility of using meson decay constants
describing the coupling of the $I=1/2$ and $I=1$ scalar mesons to
(pointlike) flavor-non-diagonal scalar densities,
as a means of further clarifying the nature of the isovector
scalar spectrum.
The basic idea is the same as that underlying attempts to make use of
the processes $f_0 ,\ a_0\rightarrow\gamma\gamma$
and $\phi\rightarrow f_0\gamma ,\ a_0\gamma$, namely that
the spatial extent of a loosely bound $K\bar{K}$ molecule
is significantly larger than that of a conventional $q\bar{q}$ meson,
which is, in turn, significantly larger than that of the
very compact Gribov minions\cite{gribov}.  The scalar densities,
being pointlike, provide a direct probe
of such scale differences,
and one which is, moreover, less
prone to obscuration by intervening dynamics.
In other channels, analogous decay constants provide useful information
on the classification of states.  For example, the fact that,
using recent ALEPH data, one has\cite{ALEPH}
\begin{equation}
g_{K^{*-}}=1.1\, g_{\rho^-}\simeq g_{\rho^-}\ ,
\label{vecdecayconstants}
\end{equation}
(where $g_{V^{ab}}$, with $a,b=u,d,s$, is defined by
$\langle 0\vert J^{ab}_\mu\vert V^{ab}\rangle \equiv g_{V^{ab}}\epsilon_\mu$,
with $J^{ab}_\mu = \bar{q_a}\gamma_\mu q_b$,
$V^{ab}$ the corresponding
vector meson, and $\epsilon_\mu$ the vector
meson polarization vector) supports the assignment
of the $\rho$ and $K^*$
to the same $SU(3)_F$ multiplet.  The same-multiplet assignment of
the $\pi$ and $K$ is similarly supported by the result
\begin{equation}
f_K=1.2\, f_\pi\simeq f_\pi\ ,
\label{psdecayconstants}
\end{equation}
while the fact that
\begin{equation}
{g^{EM}_\omega}/{g^{EM}_{\rho^0}}\simeq 1/3\ ,
\label{EMdecayconstants}
\end{equation}
(where $g^{EM}_\omega$, $g^{EM}_{\rho^0}$ are the electromagnetic
decay constants of the $\omega$ and $\rho^0$), rather than the value
$\simeq 1/\sqrt{3}$ (expected if the $\omega$ is the flavor $8$
member of a vector meson octet), gives direct evidence for ideal mixing
in the vector meson nonet.

Expectations for the relative sizes of the scalar decay 
constants in the various scenarios are obvious:  they should
be small for a weakly bound $K\bar{K}$ molecule, large for the
very compact ($\sim .1-.2\ {\rm fm}$) 
Gribov minions and, for conventional $q\bar{q}$ states,
should obey approximate $SU(3)_F$ relations among the 
the decay constants of different
members of the same multiplet.   In the four-quark picture, since,
in the bag model, a four-quark state is larger than a two-quark
state and, in addition, the hidden strange pair would
have to be annihilated,
one would also expect the decay constant to be suppressed.
We will show below that
experimental data allows us to determine the
scalar decay constant of the $K_0^*(1430)$,
while a QCD sum rule analysis fixes the analogous 
$a_0(980)$ and $a_0(1450)$ decay constants.  We will then show that
(1) the weakly bound $K\bar{K}$ molecule and minion scenarios
are ruled out by the sum rule analysis and (2) that the relations
between the various decay constants do not support the assignment
of either $a_0$ resonance as the $SU(3)_F$ partner of the 
$K_0^*(1430)$, but rather suggest a UQM-like scenario.


\section{The $K_0^*(1430)$ Scalar Decay Constant}
We define the decay constants of interest to us in this paper
as follows:
\begin{eqnarray}
\langle 0\vert \partial^\mu J^{su}_\mu \vert K_0^*(1430)\rangle
&\equiv &f_{K_0^*(1430)}m_{K_0^*(1430)}^2 \label{fk0} \\
\langle 0\vert \partial^\mu J^{du}_\mu \vert a_0\rangle
&\equiv &f^\prime_{a_0}m_{a_0}^2 \label{fpa0}
\end{eqnarray}
where, in Eq.~(\ref{fpa0}), $a_0$ refers to either of the
two $a_0$ resonances.  In QCD, one has
\begin{equation}
\partial^\mu J^{ab}_\mu = i\left(m_a-m_b\right)S^{ab}\ ,
\label{divvec}
\end{equation}
with $S^{ab}=\overline{q_a}q_b$.
Since the scalar densities, $S^{ab}$, are members
of an $SU(3)_F$ octet, $SU(3)_F$ relations are
simplified by redefining the 
$a_0$ decay constants in such a way that
a common mass factor occurs on the
LHS's of Eqs.~(\ref{fk0}) and (\ref{fpa0}), i.e.,
\begin{equation}
\left({\frac{m_s-m_u}{m_d-m_u}}\right)
\langle 0\vert \partial^\mu J^{du}_\mu \vert a_0\rangle
\equiv f_{a_0}m_{a_0}^2 \ .
\label{su3decayconstant}
\end{equation}
For an $a_0$ state lying in the same
$SU(3)_F$ multiplet as the $K_0^*(1430)$, one should then have
\begin{equation}
f_{a_0}m_{a_0}^2
\simeq f_{K_0^*(1430)}m_{K_0^*(1430)}^2\ ,
\label{su3reln}
\end{equation}
while, in the $K\bar{K}$ molecule/four-quark/minion pictures, the LHS
of Eq.~(\ref{su3reln}) should be, respectively, much smaller/smaller/larger
than the RHS.  

To make use of the expectation provided by Eq.~(\ref{su3reln}),
one first
requires the decay constant of the ``reference''
quark model state, the $K_0^*(1430)$.  This may be obtained from
the spectral function, $\rho_s (s)$, of the
correlator, $\Pi_s(q^2)$, defined by
(with $J_s=\partial^\mu J^{su}_\mu$),
\begin{equation}
\Pi_s (q^2)=i\int d^4x\, e^{iq\cdot x}
\langle 0\vert T\left( J_s(x)J_s^\dagger (0)\right)\vert 0\rangle\ ,
\label{strangecorrelator}
\end{equation}
since one has, at the $K_0^*(1430)$ peak, neglecting background,
\begin{equation}
{\frac{f^2_{K_0^*(1430)}m_{K_0^*(1430)}^3}{\pi\Gamma_{K_0^*(1430)}}}
=\rho_s (m^2_{K_0^*(1430)})\ .
\label{f0kexp}
\end{equation}
$\rho_s (s)$ may, in turn, 
be constructed from experimental
$K\pi$ phase shifts and the value
of the timelike scalar $K\pi$ form factor, $d(s)$, at some convenient
kinematic point\cite{jm,cfnp} since (1) unitarity relates
the $K\pi$ component
of the spectral function to $d(s)$ via
\begin{equation}
\left[ \rho_s (s)\right]_{(K\pi )}={\frac{3}{32\pi^2}}\sqrt{\frac{
\left( s-s_+\right)\left( s-s_-\right)}{s^2}}\, \vert d(s)\vert\ ,
\label{rhokpi}
\end{equation}
(where $s_\pm =\left( m_K\pm m_\pi\right)^2$) and, (2) $d(s)$
satisfies an Omnes relation,
\begin{equation}
d(s)=d(0)\, exp\left[ {\frac{s}{\pi}}\int_{th}^\infty\, ds^\prime
{\frac{\delta (s^\prime)}{s^\prime (s^\prime -s-i\epsilon )}}\right]\ ,
\label{omnes}
\end{equation}
where $\delta (s)$ is the phase of $d(s)$, and the
normalization $d(0)$ is known from ChPT and $K_{e3}$\cite{glmff}.
(In writing this equation, we have, as elsewhere in the
literature, ignored a possible polynomial
pre-factor; see Ref.\cite{kmms} for an empirical justification.)
Experimentally, $K\pi$ scattering is known to be
purely elastic up to
$s\sim 2.5\ {\rm GeV}^2$\cite{estabrooks,lass,dunwoodie}.
Thus, $\delta (s)$ is
identical to the $I=1/2$ $K\pi$ phase shift up to this point.
At the edge of the experimental region, $s=(1.7\ {\rm GeV})^2$,
moreover, the measured $K\pi$ phase has essentially reached
the asymptotic value, $\pi$, required by quark counting
rules for $d(s)$, allowing one to rather safely assume
$\delta (s)=\pi$ for $s>(1.7\ {\rm GeV})^2$.  With these
assumptions, $\left[ \rho_s (s)\right]_{K\pi}$
is determined by Eqs.~(\ref{rhokpi}) and (\ref{omnes}).  The self-consistency
of these assumptions has been checked by the
sum rule analysis of Ref.~\cite{kmms}.  Finally, 
the $K_0^*(1430)$ $K\pi$ branching fraction is known
to be compatible
with $100\%$\cite{dunwoodie}, so the $K\pi$ component represents
the full spectral function in the $K_0^*(1430)$ region.
(Note that Particle Data Group values
for the mass and width\cite{pdg98}
reflect
mis-transcriptions in Ref.~\cite{lass};
the corrected values are 
$m_{K_0^*(1430)}=1.412\ {\rm GeV}$ and
$\Gamma_{K_0^*(1430)}=0.294\ {\rm GeV}$\cite{dunwoodie}.
The corresponding values of the effective range parameters
appearing in the LASS parametrization of the $K\pi$ phase
are $a=2.19\ {\rm GeV}^{-1}$ and
$b=3.74\ {\rm GeV}^{-1}$\cite{dunwoodie}.)
Combining experimental data and Eqs.~(\ref{f0kexp}), (\ref{rhokpi})
and (\ref{omnes}), one obtains
\begin{equation}
f_{K_0^*(1430)} m_{K_0^*(1430)}^2 = .0842\pm .0045\ {\rm GeV}^3\ .
\label{fk0value}
\end{equation}
The errors in Eq.~(\ref{fk0value}) reflect the range of values
obtained when the $K_0^*(1430)$ resonance parameters are
varied within the errors quoted for the phase shift fit 
of Ref.~\cite{jm}, and also the difference between the values
obtained using the corrected LASS fit and the fit of Ref.~\cite{jm}.

\section{The Isovector Scalar Decay Constants}

The spectral function, $\rho (s)$, of the isovector 
scalar correlator, $\Pi (q^2)$,
receives contributions from the two $a_0$ resonances
proportional to $f^2_{a_0(980)}$ and $f^2_{a_0(1450)}$, respectively.  
This makes a QCD sum rule treatment
of $\Pi (q^2)$, in which 
the OPE of $\Pi (q^2)$
is used to fix unknown parameters
of the hadronic spectral function, such as $f_{a_0}^2$, 
an especially favorable way
to distinguish between ``large decay constant'' and
``small decay constant'' scenarios.

Previous attempts to determine $f_{a_0(980)}$ using QCD sum
rules employed the conventional 
Borel transformed (SVZ) sum rule method\cite{svz}
(see Section 7.3 of Ref.~\cite{narisonbook} 
for details).  
Borel transformation of the original dispersion relation
converts the weight in the hadronic spectral
integral into an exponentially falling one, $\exp (-s/M^2)$,
where the Borel mass, $M$, is a parameter of the transformation,
and, simultaneously, both kills subtraction constants and
provides factorial suppression of higher dimension condensate
contributions on the OPE side 
($1/\left( Q^2\right)^n\rightarrow 1/(n-1)!M^{2n}$, with $Q^2=-q^2=-s$).
Small values of $M$ suppress dependence of the transformed
spectral integral on the unknown high-$s$ portion of the
spectral function, while large values suppress dependence of
the OPE side on unknown higher dimension condensates.  Ideally,
one finds a ``stability plateau (or window)'', i.e., a
region of $M$ values for which
both suppressions are reasonably operable, and within which
extracted spectral parameters are roughly constant.  Since,
typically, the high-$s$ contribution to the transformed 
spectral integral is not negligible in the stability
window, one requires a model for this portion of the
spectral function.  The standard approach is to use the
``continuum ansatz'', or local duality approximation, for $\rho (s)$,
for all $s$ beyond some ``continuum threshold'', $s_0$.
This is known to be a crude approximation
in regions where typical resonance widths are not much greater
than typical resonance separations, and hence leads to
systematic uncertainties if continuum contributions
to the spectral integral are significant.  In the case of
the earlier sum rule analyses of $f_{a_0(980)}$, one 
sees, from Fig.~7.6 of Ref.~\cite{narisonbook}, both 
the absence of a stability plateau and a strong
sensitivity to the choice of $s_0$.  The latter observation
signals the importance of contributions from the continuum
region.  

An alternate implementation of the QCD sum rule
approach, which avoids the use of the local duality approximation,
is the finite energy sum rule (FESR) method.  Consider the
so-called ``PAC-man'' contour, which traverses both sides of 
the physical cut 
between threshold, $s_{th}$, and $s_0$, 
on the timelike $q^2=s$ axis, and
is closed by the circle of radius $s_0$
in the complex $s$-plane.  Cauchy's theorem and analyticity then ensure that
\begin{equation}
\int_{s_{th}}^{s_0}\, ds\, \rho (s) w(s)\, =\,
{\frac{-1}{2\pi i}}\, \oint_{\vert s\vert =s_0}\, ds\, w(s)\Pi (s)\ ,
\label{basicfesr}
\end{equation}
for any function $w(s)$ analytic in the region of the contour.
Typically one wishes to use spectral data and/or a spectral
ansatz on the LHS of Eq.~(\ref{basicfesr}), and the OPE on the
RHS.  The isovector scalar channel is very
well adapted to the FESR approach, as far as the
hadronic side of the sum rule is concerned, since the first
two resonances in the channel are rather well-separated.
Thus, if one chooses $s_0$ so as to include only the first two
resonance regions, a well-motivated, resonance-dominated
spectral ansatz is possible.  Since the resonance masses and widths
are known, only
the decay constants remain as unknown parameters.  
The potential problem with
the FESR approach is that those $s_0$
for which one can reliably employ a resonance-dominated
ansatz for $\rho (s)$ correspond, by definition, to scales
for which local duality is not a good approximation.  This
means, in particular, that the OPE representation for $\Pi (s)$
on the circle $\vert s\vert =s_0$ must, 
at the very least, break down over some region near the timelike
real axis.  Fortunately, an argument by Poggio, Quinn and Weinberg,
suggests that, for moderate $s_0$, this is the only
region over which this breakdown should occur\cite{pqw}.  
This is confirmed empirically
by the success of the conventional FESR treatment of hadronic
$\tau$ decays\cite{taufesr} (which involves a weight, determined by kinematics,
having a double zero at $s=s_0=m_\tau^2$), and by the investigation
of Ref.~\cite{kmfesr}, which shows that FESR's involving
weights $w(s)$ having either a single or double zero at
$s=s_0$ are all extremely well satisfied in the isovector
vector channel (where one can use the experimentally determined
spectral function\cite{ALEPH,OPAL}), even at scales, $s_0$, significantly
below $m_\tau^2$.  In Figure 1 we show that such ``pinch-weighted''
FESR's can be used to very accurately extract hadronic spectral
parameters.  In the Figure,
the dots (with error bars) are the experimental
ALEPH data\cite{ALEPH}; the
dashed line shows the result of using the OPE side of a 
continuous family of such sum rules to
fix the decay constants appearing in
a simple spectral ansatz consisting of an incoherent
sum of three Breit-Wigner resonances\cite{kmfesr};
and the solid line
represents a least squares
fit of the same spectral ansatz directly to the data.
Note the very accurate determination of the
$\rho (770)$ decay constant.  This determination is 
dominated by the perturbative ($D=0$)
contributions to the OPE,
and hence by the value of the running coupling, $\alpha_s$.
While a determination of a non-perturbative quantity
like $g_\rho$ in terms of a perturbative
one like $\alpha_s$ might sound implausible,
we remind the reader that this
is possible because we have used empirical non-perturbative
information (in the form of resonance masses and widths)
as input to the sum rules.  Once this information
has been input, analyticity relates 
the running coupling and the resonance decay constants
via the basic FESR relation Eq.~(\ref{basicfesr}).

In light of above discussion, we chose to study
the isovector scalar channel using the pinch-weighted
FESR method.
For the spectral 
function we employ a sum of 
$a_0(980)$ and $a_0(1450)$ contributions, using experimental input
for the masses and widths.  $s_0$ is then restricted
to lie at or below the upper edge of the second resonance region
(actually, somewhat higher, since the zero at $s=s_0$ means
that the region just below $s_0$ in the spectral integral
has negligible weight).  We take the maximum
value of $s_0$ to be $3.0\ {\rm GeV}^2$.  

On the OPE
side, the expressions for the dimension $D=0,2,4$ and $6$
contributions are given in
Refs.~\cite{jm,cps}, and will not be reproduced here.
The $D=0$ (perturbative) contribution is known to four-loop
order, and is determined by the running light quark
masses and coupling.  The light quark mass scale is set by
any one of $m_u, m_d, m_s$, since the mass ratios
are determined by ChPT\cite{leutwylerqmasses}.  
We employ, as our basic input, 
the value $m_s(2 \ {\rm GeV})=115\pm 8\ {\rm MeV}$
(quoted in the $\overline{MS}$ scheme) determined in Ref.~\cite{kmms}.
{\it Ratios} of decay
constants are, of course, independent of this choice.
As input for the running coupling we take 
$\alpha_s(m_\tau^2)=0.334\pm 0.022$\cite{ALEPH,OPAL}.  The masses and
couplings at other scales are obtained from the exact solutions
of the RG equations generated using the four-loop truncated versions
of the $\beta$\cite{beta4} and $\gamma$\cite{gamma4} functions.
As input for the higher dimension condensate contributions, we 
employ conventional values:  
$\langle \alpha_s G^2\rangle = 0.07\pm .01\ {\rm GeV}^4$\cite{narison97},
$\left( m_u+m_d\right)\langle \bar{u}u\rangle =-f_\pi^2 m_\pi^2$, and
$\langle g\bar{q}\sigma Fq\rangle
=\left( 0.8\pm 0.2\ {\rm GeV}^2\right)\langle \bar{q} q\rangle$\cite{jm}.
The four-quark $D=6$ condensate terms are taken to have their
vacuum saturation values, modified by an overall multiplicative 
factor, $\rho_{VSA}$.  To be conservative, we consider
the range $\rho_{VSA}=5\pm 5$, allowing, therefore,
up to an order of magnitude violation of vacuum
saturation.  Integrals around the circle $\vert s\vert =s_0$
on the OPE side of Eq.~(\ref{basicfesr}) are all performed using the
contour improvement prescription of LeDiberber and Pich\cite{pld},
which is known to improve convergence, and reduce residual
scale dependence of the truncated perturbative series.  Since
the convergence of the integrated $D=0$ contributions worsens
as $s_0$ is lowered, we have chosen $s_0=2.4\ {\rm GeV}^2$ as
a minimum value for our analysis.  Ideally, to improve convergence,
one would like
to know the mass and width of the third $a_0$ resonance,
and then work at larger scales $s_0$, 
but this is not possible at present.
We discuss our estimate of the resulting truncation errors below.

The last point in need of discussion is the treatment of
instanton effects.  These effects can be important in scalar
and pseudoscalar channels, especially at lower scales such
as those we have been forced to work at here.
The effect of instantons on $\Pi (s)$
is known exactly only in the approximation in
which one treats the single instanton
configuration in the background of the perturbative vacuum\cite{bbb,np}.
It is known, however, that quark and gluon condensates
strongly affect the density for large scale
instantons\cite{svzinstantons}.  Since
contributions in
Refs.~\cite{bbb,np}, 
from instantons large enough to be
subject to this effect are known to be significant\cite{bbb},
the exact results are of only schematic use
(as pointed out by the authors themselves).  An alternate
representation of instanton
effects is provided by the instanton liquid model, in which the
effective instanton density is assumed to be sharply
peaked around a single effective average size\cite{instantonliquid}.
Although this sharp peaking leads to a slower fall-off with $q^2$ than would
be obtained using a broader distribution of sizes\cite{chibisov},
the model has the advantage of being phenomenologically
constrained\cite{instantonliquid}.  
We have, therefore, employed the instanton liquid model.
Ref.~\cite{biggroup} gives the form of the corresponding
contributions to FESR's with $w(s)=s^k$.  In
view of the the phenomenological nature of the model, 
we chose,
to be safe, 
to restrict ourselves to weights $w(s)$ for which
instanton effects are not large.  In order that this suppression
not be specific to the 
particular $q^2$-dependence
of the instanton liquid model, we further restrict
ourselves to those weights for which an evaluation
using the much more strongly $q^2$-dependent form obtained
in Ref.~\cite{np} are also small.  This turns out to restrict us
to weights of the form
\begin{equation}
w(s)=\left( 1-{\frac{s}{s_0}}\right) \left( A -{\frac{s}{s_0}}\right)\ ,
\label{weightform}
\end{equation}
with the parameter $A$ lying in the range
$\simeq 2\pm 1$.  The suppression of the
incompletely known instanton contributions is optimal for $A=2$, and
we display all results below employing this value.  (Note that
the weight $w(s)$ has been
constructed so as to have a zero at $s=s_0$, which is required in
order that the resulting FESR's be reliable at scales such as that
considered here\cite{kmfesr}.)
We will use the difference between the results obtained using
the two different instanton implementations as a 
(hopefully conservative) measure
of the uncertainty associated with our use of the instanton
liquid model.

Fitting the $a_0(980)$ and $a_0(1450)$ decay constants by matching
the hadronic and OPE sides of the FESR above in the fit window
$2.4\ {\rm GeV}^2<s_0<3.0\ {\rm GeV}^2$, we then obtain, 
adding errors from all sources in quadrature,
\begin{eqnarray}
f_{a_0(980)}m_{a_0(980)}^2&=&0.0447\pm 0.0085\ {\rm GeV}^3 
\label{a980fm2}\\
f_{a_0(1450)}m_{a_0(1450)}^2&=&0.0647\pm 0.0123\ {\rm GeV}^3\ .
\label{a1450fm2}\end{eqnarray}
Uncertainties due to those on the resonance masses and widths 
are completely negligible; the errors, therefore, reflect
uncertainties in the input to
the OPE side of
the sum rules.  The major sources of error on
$f_{a_0(980)}m_{a_0(980)}^2$ are as follows: (1) from that
on $\alpha_s(m_\tau^2)$:  $\pm 0.0068\ {\rm GeV}^3$, (2) from
that on the overall mass scale, $\left( m_s-m_u\right)^2$:
$\pm 0.0032\ {\rm GeV}^3$, (3) from that on $\rho_{VSA}$:
$\pm 0.0023\ {\rm GeV}^3$ and (4) due to truncation
of the perturbative series:  $\pm 0.0029\ {\rm GeV}^3$.
Those on $f_{a_0(1450)}m_{a_0(1450)}^2$ are, similarly: (1) from that
on $\alpha_s(m_\tau^2)$:  $\pm 0.0112\ {\rm GeV}^3$, (2) from
that on the overall mass scale:
$\pm 0.0046\ {\rm GeV}^3$, (3) from that on $\rho_{VSA}$:
$\pm 0.0015\ {\rm GeV}^3$ and (4) due to truncation
of the perturbative series:  $\pm 0.0011\ {\rm GeV}^3$.
If one employs, instead of the instanton liquid model, the
results of Ref.~\cite{np},
$f_{a_0(980)}m_{a_0(980)}^2$ is increased by $0.0044\ {\rm GeV}^3$
and $f_{a_0(1450)}m_{a_0(1450)}^2$ by $0.0023\ {\rm GeV}^3$.
We remind the reader that the instanton liquid model is subject
to phenomenological constraints, while the
results of Ref.~\cite{np} are not.  

One final source of uncertainty, not included in the errors
quoted above, is relevant in the case of the
$a_0(1450)$.  Because the $a_0(980)$ is well separated from
subsequent resonances in the channel, its extracted decay
constant is stable with respect to assumptions about the
behavior of the spectral function beyond $3\ {\rm GeV}^2$. This is,
however, not necessarily true for the $a_0(1450)$, which might,
for example, overlap to some extent, and hence interfere 
with, the next higher resonance.
Even were this interference to be
incoherent, there could still be a non-trivial contribution
from the tail of the next $a_0$ under the $a_0(1450)$.
This would lead to an overestimate of the $a_0(1450)$ decay
constant.  We cannot meaningfully investigate this uncertainty
since we do not know the width, or location, of the next
$a_0$ resonance.  In order to get a feel for the resulting
uncertainty, however, we
have investigated the effect of 
including a third resonance in the spectral ansatz, taking, for
illustration, its mass and width to be $1.9\ {\rm GeV}$ and
$0.4\ {\rm GeV}$, respectively.  Assuming the contributions add
incoherently, one finds that the $a_0(1450)$ decay constant
is decreased by $\sim 40\%$.  That level of uncertainty in
the $a_0(1450)$ decay constant is, therefore, unavoidable
without further experimental input.  In contrast, the $a_0(980)$
decay constant is, as expected, essentially unaffected
by the use of the three-resonance ansatz.

\section{Discussion}
For results obtained using QCD sum rules to be considered
reliable, it is necessary that the form of
the spectral ansatz employed be physically sensible.
Even the most ridiculous spectral ansatz will have some
choice of parameters which
``optimizes'' the match between the OPE
and hadronic sides.  Of course, if the ansatz is not a good one,
this ``optimal'' match will be poor.  While at the scales we
have employed, resonance dominance seems a very safe assumption,
it is worthwhile checking this statement.  This is done in
Figure 2.  The Figure displays the results for the $A=2$
pinch-weighted FESR discussed above, both for the two-resonance
ansatz on which the results above are based, and for two other
ans\"atze, which serve as the basis for further discussions below.

In the Figure, the dashed-dotted line represents the OPE side of the
sum rule, the dotted line the optimized match
of the hadronic to the OPE side, obtained by adjusting
the two $a_0$ decay constants.  The agreement is obviously
excellent.  The solid line, in contrast, represents the 
match obtained when one forces $f^2_{a_0(980)}\simeq 0$ by
hand (as would be expected for a loosely-bound $K\bar{K}$
molecule) and adjusts $f^2_{a_0(1450)}$ to optimize the match.
The resulting fit
is very poor, ruling out the loosely-bound
$K\bar{K}$ interpretation of the $a_0(980)$.  Similarly, the
fact that no good match exists with $f_{a_0(980)}$ much
larger than $f_{K_0^*(1430)}$ rules out the minion interpretation.
A sceptical reader might object that the
poor fit between the OPE and 
hadronic sides in the $a_0(1450)$-only spectral ansatz might be cured,
not only by the addition of a narrow $a_0$ contribution, but also by
the inclusion of a broader non-resonant background,
even though such a possibility appears somewhat perverse, in view of the
extremely good match between the dashed-dotted and dotted curves.
The dashed line in the Figure demonstrates that this possibility
is also ruled out, for reasons we will now explain.

Although we do not know, in detail, what to expect for the background
$\pi\eta$ and $K\bar{K}$ contributions to the isovector scalar
spectral function below the $a_0(1450)$ region, 
it is rather easy to make sensible rough estimates.
Indeed, once one knows the scalar form factors describing the
couplings of the scalar densities to the $\pi\eta$ and $K\bar{K}$ states,
one obtains the corresponding contributions to the spectral
function by unitarity, as in Eq.~(\ref{rhokpi}).  Near threshold,
the timelike form factors may be trivially computed using ChPT.
If one performs this exercise for the $\pi\eta$ contribution,
using the tree-level ChPT expression for the timelike form-factor,
the optimized background-plus-$a_0(1450)$ fit is very close to
that of the solid line in Figure 2.  The dashed line is the
result of multiplying this background contribution by a factor
of $5$, and then adjusting $f_{a_0(1450)}$ so as to optimize the
match between the OPE and hadronic sides (but still with no
explicit $a_0(980)$ contribution).  Again one sees that the fit
is very poor, demonstrating that it is a narrow state, and not
a broad background which is required in the spectrum at low $s$.

We, thus, conclude that the results obtained above
are, indeed, reliable, subject to the quoted errors and the caveats
regarding the $a_0(1450)$ decay constant already discussed.
We see that (1) the products $f_{a_0}m_{a_0}^2$
for the two $a_0$ resonances
are comparable and (2) that both are somewhat
smaller than the corresponding product for the $K_0^*(1430)$.
In addition to ruling out the $K\bar{K}$ molecule and minion
pictures, these results suggest a UQM-like
scenario.  One ambiguity which is unavoidable concerns the
way in which one interprets the expected $SU(3)_F$ relations
amongst the decay constants for the states lying in the same
multiplet as the $K_0^*(1430)$.  In the $SU(3)_F$ limit, of
course, all states in the same multiplet would have the same
mass, so whether one compared the values of $fm^2$ or the values
of $f$ would make no difference.  In attempting to decide
whether or not the $a_0(980)$ should be interpreted as the
$I=1$ partner of the $K_0^*(1430)$, however, this ambiguity
plays a potentially significant role:  if we compare decay
constants, then we have
\begin{equation}
{\frac{f_{a_0(980)}}{f_{K_0^*(1430)}}}=1.1
\label{fratio}
\end{equation}
whereas, if we compare the products of the decay constants
and the squared masses, we have
\begin{equation}
{\frac{f_{a_0(980)}m^2_{a_0(980)}}
{f_{K_0^*(1430)}m^2_{K_0^*(1430)}}}=0.53\ .
\label{fm2ratio}
\end{equation}
Since, in other channels, it is the strange state which has the
larger decay constant, however, neither of these results, in fact,
corresponds to what one might expect based on the pattern from 
other meson nonets.

In conclusion, we have determined the scalar decay constants
of the $a_0(980)$, $a_0(1450)$ and $K_0^*(1430)$ mesons.
The relations between them suggest a UQM-like scenario for
the isovector scalar states.  At present, we do cannot be
certain that the results rule out the four-quark interpretation
of the $a_0(980)$, though it appears likely that one
would expect a much smaller value for $f_{a_0(980)}$
in this scenario.  A calculation of the 
$f_{a_0(980)}$ in the four-quark
picture would thus be highly desirable.

\acknowledgements
The author would like
to thank M. Pennington for a number of very useful conversations,
and for bringing the existence of the transcription errors
in the published version of the LASS parametrization to his
attention; W. Dunwoodie for confirming the
revised values of the $K_0^*(1430)$ fit parameters; and
A. H\"ocker for a number of discussions concerning the ALEPH spectral
data and their interpretation.
The ongoing support of the Natural Sciences and
Engineering Research Council of Canada, and the hospitality of the
Special Research Centre for the Subatomic Structure of Matter at the
University of Adelaide, where this work was performed, are also
gratefully acknowledged.

\vskip .5in\noindent
 \begin{figure} [htb]
\centering{\
\psfig{figure=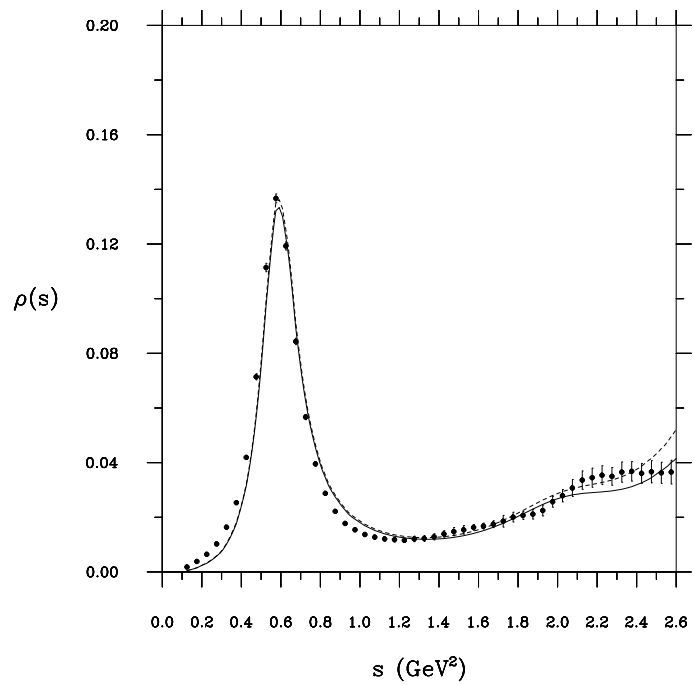,height=10.cm}}
\vskip .3in         
\caption{Comparison of data, the representation
generated by fitting to the OPE, and the least squares fit to data, 
for the isovector vector spectral function.  See the text for
details.}
\label{figone} 
\end{figure}

\vskip .5in\noindent
 \begin{figure} [htb]
\centering{\
\psfig{figure=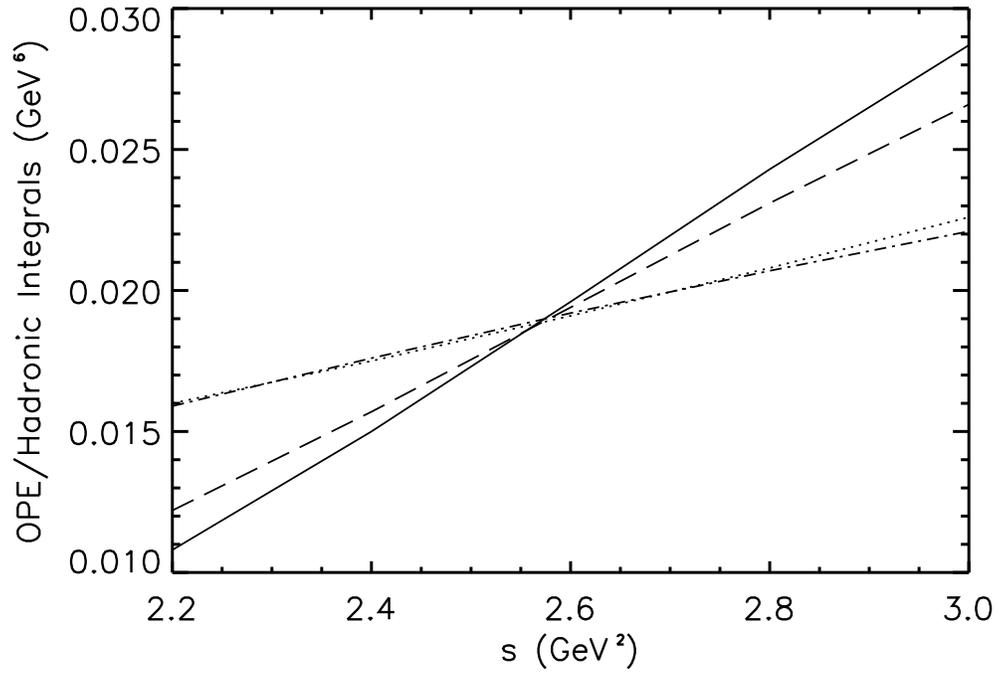,height=10.cm}}
\vskip .3in         
\caption{The $A=2$ pinch-weighted FESR.  The curves correspond to
the OPE side of the sum rule and the hadronic sides
corresponding to the three different spectral ans\"atze described in
the text.}
\label{figtwo} 
\end{figure}

\end{document}